\documentclass[aps,twocolumn,amsfonts,amssymb,longbibliography,superscriptaddress]{revtex4-1}
\usepackage{graphicx}
\usepackage{multirow}
\usepackage{epstopdf} 
\usepackage{dcolumn}
\usepackage{bm}
\usepackage{bbm,bbold}
\usepackage{color}
\usepackage{xcolor, soul}
\sethlcolor{green}
\usepackage{amsfonts}
\usepackage{amsmath}
\usepackage{mdframed}
\usepackage[normalem]{ulem}
\usepackage{mathrsfs}   
\usepackage[none]{hyphenat}
\usepackage{subfigure}
\usepackage{float}
\usepackage{physics}
\usepackage[colorlinks=true,citecolor=blue]{hyperref}
\hypersetup{colorlinks=true,citecolor=blue,linkcolor=blue,urlcolor=blue}

\begin{document}
	\title{Reshaping the anomalous Hall response in tilted 3D system with disorder correction}	
	\author{Vivek Pandey}
    \email{vivek_pandey@srmap.edu.in}
	\affiliation{Department of Physics, School of Engineering and Sciences, SRM University AP, Amaravati, 522240, India}
 	\author{Pankaj Bhalla}
	\email{pankaj.b@srmap.edu.in}
 \affiliation{Department of Physics, School of Engineering and Sciences, SRM University AP, Amaravati, 522240, India}
 \affiliation{Centre for Computational and Integrative Sciences, SRM University AP, Amaravati, 522240, India}
    
\date{\today}

\begin{abstract}
The anomalous Hall conductivity in the nodal line semimetals (NLSMs) due to the presence of a symmetry-protected nodal ring adds complexity in the investigation of their transport properties.
By employing quantum kinetic theory and considering the weak disorder limit, we analyze the intraband and interband parts of anomalous Hall conductivity in the tilted 3D Dirac NLSMs. 
Our findings reveal that the net anomalous response is mainly contributed by the interband part. Further, the latter part gives non zero results by breaking inversion symmetry via tilt. 
We observe that the competition between the tilt and the chemical potential emerges kinks at distinct characteristic frequencies in the intrinsic interband part of the anomalous conductivity. On the other hand, the disorder driven interband component of the conductivity exhibits a prominent peak at low chemical potential, followed by a sign change. Notably, the disorder or extrinsic contribution to the response dominates over the intrinsic interband contribution, making it a crucial factor for the study of the overall response of a three-dimensional system.  
\end{abstract}

\maketitle

\section{Introduction}
The generation of transverse voltage in a material in the absence of the magnetic field by electrical means gives birth to a phenomenon known as the anomalous Hall effect. Recently, anomalous Hall conductivity (AHC) has become a focal point of present-day research in two and three-dimensional topological materials such as Dirac semimetals (DSMs)~\cite{Nandy_prb2019, Burkov_prb2018}, Weyl semimetals (WSMs)~\cite{Burkov_prl2014, Shekhar_pnas2018, Mukherjee_jpcm2018, Burkov_prb2018, Chatterjee_jpcm2023}, and Nodal line semimetals (NLSMs)~\cite{Ahn_prl2017, Wang_prb2021, Flores-Calderón_EL2023, Chatterjee_jpcm2023}. The origin of AHC comes via two channels, namely the band structure topology and scattering events. The contributions to AHC led by the geometrical property of the electron wave function such as quantum metric are termed intrinsic. While the contributions from random disorder potential and phonons to AHC are known as extrinsic~\cite{Nagaosa_rmp2010, Culcer_ecmp2024}. Furthermore, the presence of the spin-orbit coupling (SOC) is also crucial to generate AHC in symmetry-protected topological materials~\cite{Flores-Calderón_EL2023}. On the other hand,  in the absence of SOC the AHC can be generated by breaking the time reversal ($\mathcal{T}$) or inversion ($\mathcal{P}$) or combined $\mathcal{PT}$ symmetry~\cite{Wang_prb2021}. 
\begin{figure}[t]
    \centering
    \includegraphics[width=7.5 cm]{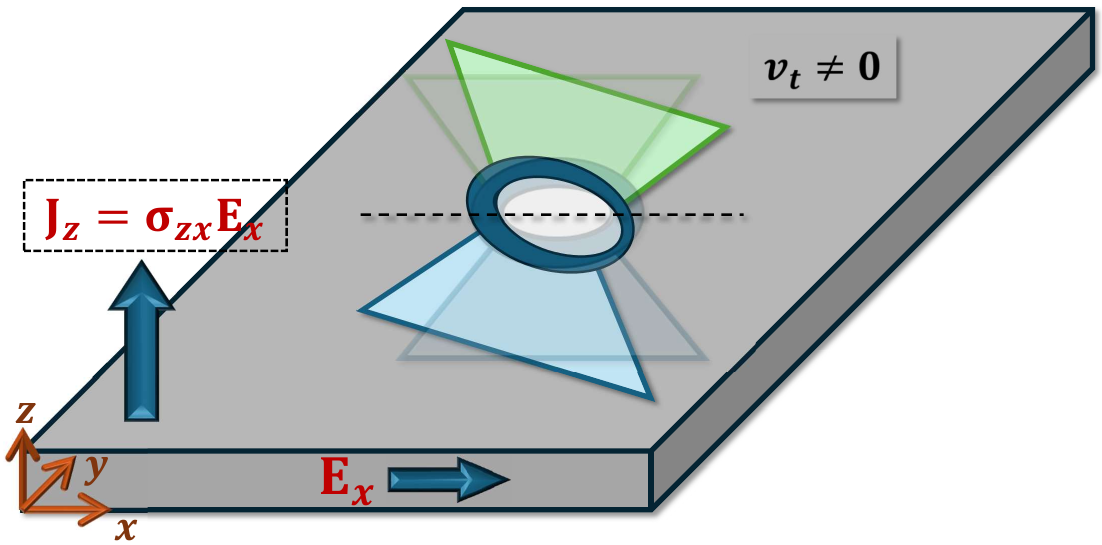}
    \caption{A schematic diagram shows the emergence of the anomalous Hall current $J_z$ with the application of the external field along $x$-direction, $E_x$ for a tilted DNLSM where the tilt $v_t$ is taken along $x$-direction. The resulting response is the artifact of the intrinsic (field-driven) and extrinsic (scattering-driven) contributions.} 
    \label{fig:1}
\end{figure}
Unlike Dirac~\cite{Wang_prb2013, liu_nm2014, Borisenko_prl2014, yi_sr2014, liu_science2014, xu_science2015} and Weyl semimetals~\cite{Lv_prx2015, Weng_prx2015, xu_nc2016, Belopolski_prl2016, Souma_prb2016, Xu_sa2015}, the electronic band structure of NLSMs shows closed loops or continuous lines in the momentum space due to band crossing instead of discrete points. Further, this band crossing is protected by symmetries such as $\mathcal{P}$ or $\mathcal{T}$ and leads to intriguing optical and electronic properties~\cite{Burkov_prb2011, Bian_nature2016, Shuo_apx2018, Bian_prb2016, Fang_prb2015, Yu_prl2015, Xie_APLM2015, Kim_prl2015, Ekahana_njp2017, Feng_prm2018, schoop_nc2016, Neupane_prb2016, Hu_prl2016, Takane_prb2017, Chen_prb2017, Wang_prb2017, Liang_prb2016}.
Further, the NLSMs on the basis of band degeneracy can be segregated into two parts: Dirac nodal line semimetals (DNLSMs) and other is Weyl nodal line semimetals (WNLSMs)~\cite{Guo_mf2022, Fang_prb2015, Kim_prl2015, Zhang_prb2017, Barati_prb2017, Weng_prb2015, Mullen_prl2015, Pandey_prb2024}. The unique features of DNLSMs, such as the flat drumhead surface states make these materials more exciting. A few examples of $\mathcal{PT}$ symmetric DNLSMs are Ca$_3$P$_2$, Cu$_3$N, PtSn$_4$ and ZrSiS~\cite{Xie_APLM2015, Chan_prb2016, Kim_prl2015, Fu_sa2019,Lin_nl2024}. On introducing the mass term (gap), $\mathcal{PT}$ symmetry of the system breaks down and the gap opens up around the Dirac point. Further, the mass term stems from the effect of disorder, circularly polarized light, pressure, uniaxial strain or external electric field~\cite{chiba_prb2017, Kot_prb2020, chen_prb2018, Wang_prb2021, Rendy_jap2021, du_nrp2021, Flores-Calderón_EL2023}. 

Moreover, NLSMs due to the presence of the interesting band crossing, host exciting transport properties, including the anomalous Hall effect~\cite{Burkov_prb2018, Wang_prb2021}, planar Hall effect~\cite{Burkov_prb2018, Li_prb2023}, quantum anomalies (chiral anomaly)~\cite{Burkov_prb2018}, and parity anomaly in DNLSMs~\cite{Rui_prb2018}.
In addition, the longitudinal conductivities, including those of massive DNLSMs, have been studied extensively in both DC and AC regimes (including intraband and interband parts)~\cite{Barati_prb2017, Pandey_prb2024, Carbotte_jpcm2017, Mukherjee_prb2017, Barati_prb2017}. 

Furthermore, to observe a finite anomalous Hall effect (AHE) in a 2D system, it is necessary to break $\mathcal{T}$ symmetry of the system~\cite{Nagaosa_rmp2010}. 
However, the situation does not remain the same for 3D NLSMs which possess a nodal ring. Such systems in the presence of $\mathcal{PT}$ breaking or more specifically in the presence of $\mathcal{T}$ symmetry breaking mass term and absence of SOC generate transverse current at each point in the nodal ring. Nevertheless, because of the existence of inversion ($\mathcal{P}$) symmetry points in the nodal ring, the net Hall current vanishes. This occurs due to the opposite signs of the transverse current at symmetrical points. 
Therefore, the presence of $\mathcal{P}$ symmetry breaking tilt term is crucial to get the non-zero anomalous Hall current in the 3D NLSMs~\cite{Wang_prb2021}.
\begin{figure*}[htp]
    \centering
    \includegraphics[width=15 cm]{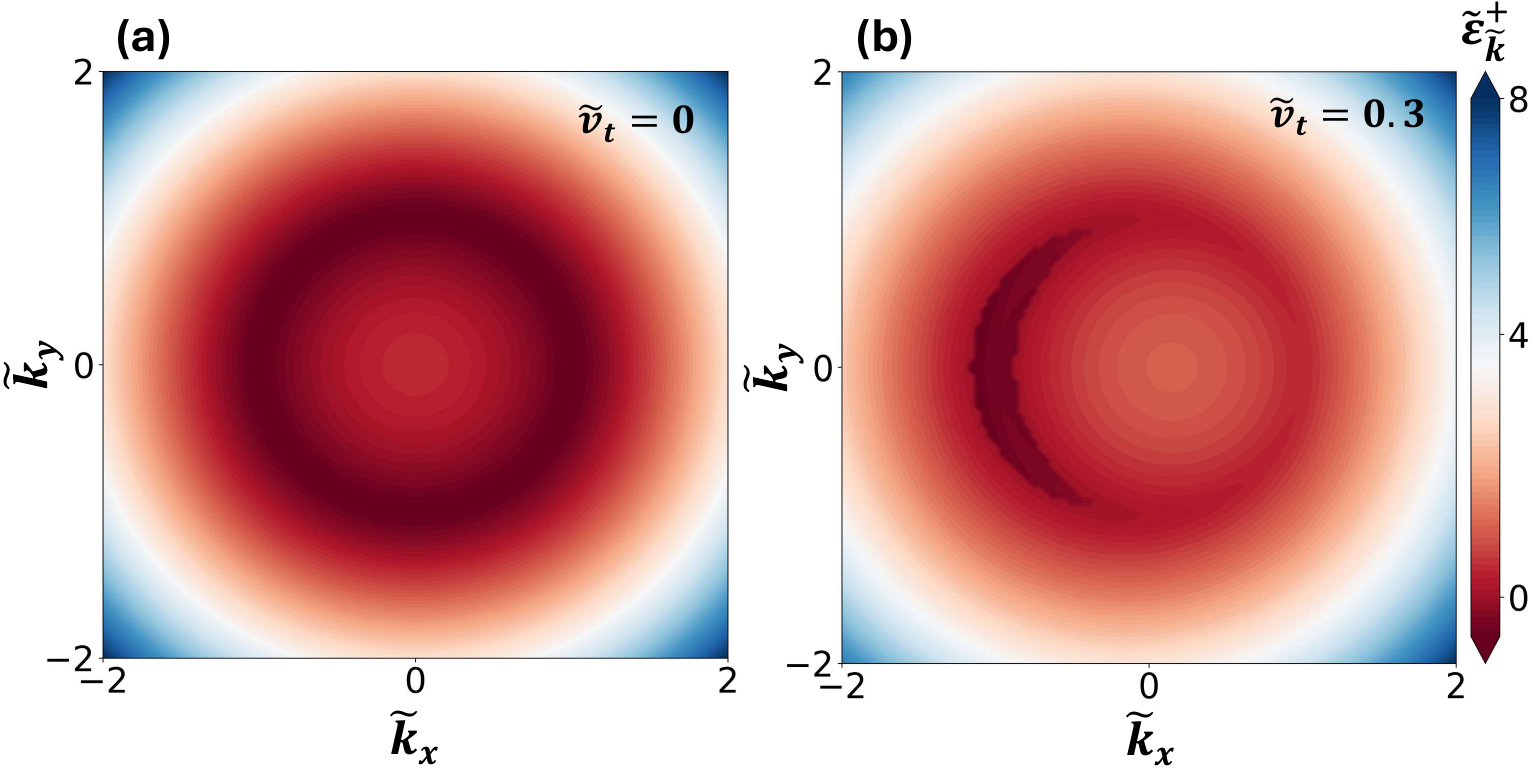}
    \caption{Contour plot for the energy dispersion $\tilde{\varepsilon}_{\bm k}^{+} = \varepsilon_{\bm k}^{+}/\varepsilon_0$ at $\tilde{k}_z=0.1$ for tilted DNLSM at tilt values (a) $\tilde{v}_t = 0$ and (b) $\tilde{v}_t =0.3$. The color bar here refers to the magnitude of the energy and the wave vectors $\tilde{k}_x = k_x/k_0$, $\tilde{k}_y = k_y/k_0$.} 
    \label{fig:Dispersion}
\end{figure*}

The presence of the tilt in DNLSMs generates peculiar properties, as it can introduce a competition between the energy dispersion and the chemical potential~\cite{Ahn_prl2017}. This interplay leads to transitions in the total response of the system at distinct characteristic frequencies~\cite{Wang_prb2021}. It is apparent that the tilt in NLSMs can be incorporated by breaking inversion or time-reversal symmetry, and have been experimentally observed in $\text{ZrSiSe}$, $\text{ZrSiS}$ and $\text{HfSiS}$~\cite{Chen_prb2017, shao_np2020}.
To elaborate further, significant progress has been made in understanding the effect of tilted NLSMs. Using the low-energy model for tilted NLSMs, the longitudinal AC response, including both intra- and interband regimes~\cite{Ahn_prl2017}, has been discussed. The study highlights the geometry of the Fermi surface under varying tilt and the chemical potential. 
Additionally, Wang et al.~\cite{Wang_prb2021} explored the anomalous Hall response in $\mathcal{PT}$ symmetry broken tilted NLSMs by applying the Kubo formalism. Further, they have demonstrated that the competition between the tilt term and the chemical potential shows kinks in the distinct characteristic frequencies. However, they have not considered the scattering effect or disorder effect which is a crucial player for total AHC. In Ref.~\cite{Pandey_prb2024} the authors calculate the longitudinal interband response of the Dirac nodal line semimetals (DNLSMs), considering both the intrinsic (field driven) and extrinsic (scattering driven) contributions. However, the anomalous Hall response in their system remains restricted due to symmetry arguments. These studies offer valuable insights into the directional dependence and symmetry-breaking effects on the transport properties of topological semimetals. However, the emphasis is given to the intrinsic part of AHC and the extrinsic part to AHC has not been explored yet which will play a significant role to have overall anomalous Hall response of DNLSMs. This motivates further investigation of the 
total response of the tilted system stemming from the combination of the intrinsic and extrinsic parts. 
To get deeper insights of the total response of DNLSMs, we consider the following aspects which distinguish our study from the previous works: (1) Both intrinsic and extrinsic contributions have been taken into account and treated on equal footing, 
(2) the importance of the inclusion of inversion symmetry breaking terms to have both finite intrinsic and extrinsic contributions to AHC, 
(3) the tunability of response with the chemical potential, 
(4) the effect of the intraband part of the density matrix on the interband contribution, which gives the extrinsic component of the interband part of conductivity, (5) the present framework encapsulates the Fermi sea and Fermi surface contributions explicitly.
In Fig.~\ref{fig:1}, we show the schematic picture of the generation of anomalous current along the $z$-direction in tilted DNLSMs by introducing tilt along the $x$-direction ($v_t\hat{x}$) and applying an external field in the $x$-direction. It is important to note here that both the intrinsic and extrinsic contributions of the interband part of the anomalous Hall conductivity are non-zero only due to the inversion symmetry broken tilt term. Furthermore, the intraband and extrinsic contribution of the interband part of the anomalous conductivity depends on the derivative of the Fermi Dirac distribution function, thus refers as the Fermi surface effect, whereas the intrinsic contribution to the interband part of anomalous conductivity is proportional to the Fermi distribution function and denoted as the Fermi sea effect.  

Our paper is organized as follows. In Section~\ref{sec:model and method}, we discuss the model~\ref{subsec:model} where we have shown the Hamiltonian for the tilted DNLSMs, followed by the Method subsection~\ref{subsection:method} provides the recipe for the quantum kinetic approach to obtain the total contribution of conductivity~\ref{subsec: total conductivity}, which is the combination of the intraband and interband part (combination of intrinsic and extrinsic contributions) to the conductivity and symmetry regarding the model is analyzed in subsection~\ref{subsec:symmetry}. In Section~\ref{sec: results}, we present numerical results for tilted DNLSMs followed by the validity of the results~\ref{subsec:validity} and experimental relevance along with numerical estimation using the real material parameters~\ref{subsec:Experimental relevance}. Finally, in Section~\ref{sec: conclusion}, we have summarized our work and provided future directions.

\section{Model and Method}\label{sec:model and method}
\subsection{Model}\label{subsec:model}
In this study, we consider tilted DNLSMs as a system, described by the low-energy Hamiltonian~\cite{Wang_prb2021}, 
\begin{equation}\label{eqn:hamitonian}
    \mathcal{H}({\bm k}) =  \frac{\hbar^{2}k_0^2} {2m}\space\bigg[(\mathcal{\tilde{K}}-1) \space\sigma_x +  \gamma \tilde{k}_z \space\sigma_y + \tilde{M} \space \sigma_z + \tilde{{v}}_t \tilde{k}_x \space \sigma_0 \bigg]. 
\end{equation}
Here, $\tilde{\mathcal{K}} =\mathcal{K}/k_0$ is the normalized radial wave vector that has $\mathcal{K} = {\sqrt{k_x^2 + k_y^2}}$ and $k_0$ as the radius of the nodal ring, $ \tilde{k}_i=k_i/k_0$, $k_i \equiv (k_x, k_y, k_z)$ refers to the wave vector and $\gamma=2 m v_z/\hbar k_0$, here $v_i \equiv (v_x,v_y,v_z)$ is the velocity of an electron in $i^{th}$ direction, $m$ corresponds to the mass of the electron. Furthermore, $\sigma_i \equiv
(\sigma_x, \sigma_y, \sigma_z)$ refers to the Pauli matrix in a pseudo spin basis and $\tilde{M}=M/\varepsilon_0$, where $M$ corresponds to the gap (mass term) and $\varepsilon_0=\hbar^2 k_0^2/(2m)$ is the energy corresponding to the nodal ring. The tilt parameter $\tilde{v}_t = 2m v_t/\hbar k_0$, where $v_t$ is the tilt in the system, taken along the $x$ direction, as the tilt along out of a plane ($z$ direction) has minimal effect on conductivity. 
In the absence of mass term and tilt, the fundamental symmetries ($\mathcal{T}, \mathcal{P}, \mathcal{PT}$) of the system remain preserved. However, the $\mathcal{PT}$ symmetry of the system does not remain invariant on the inclusion of the mass term in the Hamiltonian $M \sigma_z$ and the latter term opens up the gap between the valance and conduction bands, which will be intact at $M=0$. Experimentally the mass term can be originated via numerous ways like disorder, uniaxial strain, circularly polarized light, pressure, etc~\cite{chiba_prb2017, Kot_prb2020, chen_prb2018, Wang_prb2021,Rendy_jap2021, du_nrp2021, Flores-Calderón_EL2023, Pandey_prb2024}. 
Furthermore, the tilt term also breaks the inversion symmetry of the system which plays an essential parameter to generate asymmetry in the band dispersion which will affect the distribution function and the velocity component of the system, hence the finite anomalous Hall conductivity in the later calculation.
The energy eigenvalues for the Hamiltonian (Eq.~\eqref{eqn:hamitonian}) are
\begin{equation}
    \tilde{\varepsilon}_{{\bm k}}^n = \tilde{v}_t \tilde{k}_x +n  \tilde{\varepsilon}_{{\bm k}},
\end{equation}
here, $\tilde{\varepsilon}_{{\bm k}}=\sqrt{(\tilde{\mathcal{K}} - 1)^2+(\gamma \tilde{k}_z)^2 + \tilde{M}^2}$. Further we set, $ \tilde{\varepsilon}_{{\bm k}}^n = \varepsilon_{{\bm k}}^n/\varepsilon_0$, and $n$ is the band index, corresponds to the conduction band ($+$) and the valence band ($-$). The corresponding energy dispersion is shown in Fig.~\ref{fig:Dispersion} (a) and (b) at tilt parameter $\tilde{v}_t = 0$ and $\tilde{v}_t\neq 0$.
The eigenfunctions associated with the Hamiltonian are
\begin{equation} \label{eqn:estate}
    |u_{{\bm k}}^{n} \rangle = \frac{1}{\sqrt{2}} \left(
    \begin{matrix}
        -\sqrt{1 - \frac{\tilde{M}}{n \tilde{\varepsilon}_{{\bm k}}}} e^{-i\theta_{\tilde{\bm{k}}}} \\
        \sqrt{1 + \frac{\tilde{M}}{n \tilde{\varepsilon}_{{\bm k}}}}
    \end{matrix}
    \right),
\end{equation}
where $\theta_{\tilde{\bm{k}}}$ is the angle between $\tilde{\mathcal{K}}$ (i.e., $\tilde{k}_x$ and $\tilde{k}_y$ plane) and $\tilde{k}_z$-direction and is defined as $\theta_{\tilde{\bm{k}}} = \tan^{-1} \bigg[\frac{\gamma \tilde{\bm{k}}_z}{(\tilde{\mathcal{K}}-1)}\bigg]$.

\subsection{Method}\label{subsection:method}
We begin by considering the quantum Liouville equation for the single-particle density matrix $\rho$ in $k$- space that describes the time evolution of the density matrix~\cite{pottier_OUP, Culcer_prb2017, Bhalla_prb2023},
\begin{equation} \label{eqn:QLE}
    \frac{\partial \rho}{\partial t}+i \mathcal{\bm{L}} \rho = 0. 
\end{equation}
Here, $\mathcal{L}$ denotes quantum Liouville operator defined by  $\mathcal{\bm{L}}* = \frac{1}{\hbar}[\mathcal{H},*]$, where the symbol $*$ shows any operator and $[\cdot,\cdot]$ represents the commutator bracket. The total Hamiltonian of the system $\mathcal{H}= \mathcal{H}_0+\mathcal{H}_{\bm E}+U$, is the combination of the band part $\mathcal{H}_0$ (unperturbed), field part $\mathcal{H}_{\bm E}$ (perturbed), and the disorder part $U$ respectively. 
The field part $\mathcal{H}_{\bm E} =e\bm{E}$.$\hat{\bm{r}}$, originates from the interaction from the external electric field $\bm{E}$, here we have taken it as time-dependent and spatially homogeneous and $\hat{\bm{r}}$ is a vector associated with the position of the electron, and $e$ represents the electronic charge. 
It is important to highlight that we consider a weak disorder potential in our study. Here, the average of the disorder matrix element is $\langle U(\bm {r})\rangle = 0$. Further, the multiple averages of disorder matrix elements in a unit volume are $\langle U(\bm{r})U(\bm{r}')\rangle=U_0^2\delta(\bm{r}-\bm{r}')$~\cite{Culcer_prb2017}. The $U_0$ incorporates the strength of the disorder potential. Note that considering weak disorder potential, the treatment of the disorder part is done within the first-order Born approximation. 

To understand the intraband and interband dynamics it is convenient to break the density matrix into two parts,  
 \begin{align} \label{eqn:dmatrix}
     \rho_{\bm{k}} = \rho_{\bm k}^{\text{Intra}}+\rho_{\bm k}^{\text{Inter}},  
 \end{align}
where $\rho_{\bm k}^{\text{Intra}}=\rho_{0,\bm{k}}^{nn} + \mathcal{N}_{\bm E,\bm{k}}^{nn}$ is the intraband part of the density matrix and $\rho_{\bm k}^{\text{Inter}}=S_{\bm E,\bm{k}}^{np}$ refers to the interband part of the density matrix. Further, $\rho_{0,\bm{k}}^{nn}=f^0_{\bm{k}}=\big[1+e^{\beta(\varepsilon_{\bm{k}}-\mu)}\big]^{-1}$ is the equilibrium Fermi-Dirac distribution function, having $\beta = [k_BT]^{-1}$, $k_B$ denotes the Boltzmann constant, $T$ and $\mu$ represents the absolute electronic temperature and the chemical potential respectively. The terms $\mathcal{N}_{\bm E,\bm{k}}^{nn}$ and $S_{\bm E,\bm{k}}^{np}$ in Eq.~\eqref{eqn:dmatrix} are field correction terms. 
First solving Eq.~\eqref{eqn:QLE} for the intraband contribution by replacing $\rho$ with $\mathcal{N}_{\bm E,\bm{k}}$ and taking the total Hamiltonian of the system, we obtain 

\begin{equation}\label{Eq:N_E}
    \mathcal{N}_{\bm E,{\bm{k}}}^{nn}=\frac{e\space\bm{E}}{(g_{\bm k}+i\hbar\omega)} \cdot\frac{\partial f^0_{\bm{k}}}{\partial \bm{k}}.
\end{equation}
Note that here the commutator $\langle n| [H_0,\mathcal{N}_{\bm E,{\bm{k}}}]|n\rangle = 0$ and we considered $\langle n| [U,\mathcal{N}_{E,{\bm{k}}}]|n\rangle = \mathcal{N}_{\bm E,{\bm{k}}}/\tau_{\bm {k}}^n$, where $\ket{n}= e^{-i\bm{k.r}}\ket{u_{\bm{k}}^n}$ denotes the eigenfunction of the system. Moreover,  $g_{\bm k} = \frac{\hbar}{ \tau^n_{\bm k}}$ the scattering energy scale,  $ \partial/\partial {\bm {k}}$ is the partial derivative with respect to the wave vector, $\tau^n_{\bm k}$ denotes the relaxation time associated with the intraband scattering events in the presence of the short-range impurities and it is represented as $\frac{1}{\tau_{\bm k}^n}=\frac{n_i\space U_0^2}{\hbar}\int d\bm k'\space \space\delta(\varepsilon_{\bm {k}'}^n- \varepsilon_{\bm {k}}^n)$, where $n_i$ is the impurity density. Further, on solving Eq.~\eqref{eqn:QLE} for the interband part of the density matrix $S_{\bm E,{\bm k}}^{np}$ we get
\begin{equation} \label{eqn:6}
    \frac{\partial  S_{\bm E,{\bm{k}}}^{np}}{\partial t}+\frac{i}{\hbar} [ \mathcal{H}_0, S_{\bm E,{\bm{k}}}]^{np}+\mathcal{J}_{\bm{k}}^{np}[\mathcal{N}_{\bm E,{\bm k}}] + \frac{S_{\bm E,{\bm{k}}}^{np}}{\tau}={D_{\bm E,{\bm{k}}}^{np}},
\end{equation}
where $\langle n| [ \mathcal{H}_0, S_{\bm E,{\bm{k}}}]|p\rangle= S_{\bm E,{\bm{k}}} (\varepsilon_{\bm k}^{n}-\varepsilon_{\bm k}^{p}) $, the right-hand side refers to the driving term and in the band basis representation it is defined like $\hbar D_{\bm E,\bm{k}}^{np} =\hbar \langle n|[\mathcal{H}_{\bm E},\rho]\ket{p} = e\bm{E}.\big[\mathcal{D}_{\bm{k}}\rho\big]^{np}$, where $\big[\mathcal{D}_{\bm{k}}\rho\big]^{np} = \partial_{\bm {k}} \rho^{np}-i[\mathcal{R}_{\bm{k}},\rho]^{np}$ is a covariant derivative~\cite{Nagaosa_AM2017, Bhalla_prb2023}, where $\partial_{\bm k}\equiv \partial/\partial \bm k$. The Berry connection in the $\bm{k}$-space is denoted by $\mathcal{R}_{\bm{k}}^{np}=\bra{u_{\bm{k}}^n}\ket{i\partial_{\bm{k}}u_{\bm{k}}^p}$, the scattering term $\mathcal{J}_{\bm{k}}^{np}[\rho_{\bm E,{\bm k}}] = \mathcal{J}_{\bm{k}}^{np}[\mathcal{N}_{\bm E,{\bm k}}]+\mathcal{J}_{\bm{k}}^{np}[\mathcal{S}_{\bm E,{\bm k}}]$ where the first part takes into account the scattering effects via the intraband contributions which we have treated within the first order Born approximation in the later calculations and the second part to the scattering effects via interband contributions which we have considered within the relaxation time approximation as $ \mathcal{S}_{\bm E,{\bm{k}}}^{np}/\tau$  having $\tau$ represents the relaxation time associated with the interband effects and is treated as a constant~\cite{Pandey_prb2024}. The scattering term in the weak disorder limit in the band basis representation is defined in the form
\begin{align}\nonumber\label{eqn:J}
\mathcal{J}_{\bm{k}}^{np}[\mathcal{N}_{\bm E, \bm k}]=\frac{1}{\hbar^2}\sum_{\bm{k}'}\sum_{q}\bigg[\frac{U_{\bm{kk}'}^{nq}U_{\bm{k}'\bm{k}}^{qn}\space\mathcal{N}_{\bm E,\bm{k}}^{qp}}{i\big(\varepsilon_{\bm{k}'}^q-\varepsilon_{\bm{k}}^n\big)/\hbar}-\frac{U_{\bm{kk}'}^{nq}U_{\bm{k}'\bm{k}}^{np}\space\mathcal{N}_{\bm E,\bm{k}'}^{qn}}{i\big(\varepsilon_{\bm{k}'}^n-\varepsilon_{\bm{k}}^p\big)/\hbar}\\
-\frac{U_{\bm{kk}'}^{nq}U_{\bm{k}'\bm{k}}^{np}\space\mathcal{N}^{qn}_{\bm E,\bm{k}'}}{i\big(\varepsilon_{\bm{k}'}^n-\varepsilon_{\bm{k}}^q\big)/\hbar}+\frac{U_{\bm{kk}'}^{qn}U_{\bm{k}'\bm{k}}^{np}\space\mathcal{N}^{nq}_{\bm E,\bm{k}}}{i\big(\varepsilon_{\bm{k}'}^q-\varepsilon_{\bm{k}}^n\big)/\hbar}\bigg].
\end{align} 
Here, the averages of the product of two disorder matrix elements are expressed in the form $U_{\bm{kk}'}^{nq} U_{\bm{k}'\bm{k}}^{qt} = U_0^2 \langle u_{\bm{k}}^n | u_{\bm{k}'}^q \rangle \langle u_{\bm{k}'}^q | u_{\bm{k}}^t \rangle$ with $U_{\bm{kk}'}^{nq} = \langle n, \bm{k} | U | q, \bm{k}' \rangle$ as the disorder potential matrix~\cite{Yang_prb2011}.
On solving the Eq.~\eqref{eqn:6}  we obtain, 
\begin{align}\label{eqn:S_E}
    S_{\bm E,{\bm{k}}}^{np}= \frac{{D_{\bm E,{\bm{k}}}^{np}} - \mathcal{J}_{\bm{k}}^{np}[\mathcal{N}_{\bm E,{\bm{k}}}]}{g+\space i\space ({\varepsilon}^{n}_{\bm k}- {\varepsilon}^{p}_{\bm k}+ \hbar \omega)}.
\end{align}
Here $g = \frac{\hbar}{\tau}$, ${D_{\bm E,{\bm{k}}}^{np}}$ leads the field-driven (intrinsic) part and $\mathcal{J}_{\bm{k}}^{np}[\mathcal{N}_{\bm E,{\bm{k}}}]$ gives the scattering-driven (extrinsic) part to the interband contribution of the total conductivity. The detailed derivation of the above equation is provided in ~\ref{Appendix:A1}.

\subsection{Total conductivity of DNLSMs}\label{subsec: total conductivity} 
In general, the total conductivity is the sum of the intraband and interband parts of the conductivity $\sigma^{\text{Total}}=\sigma^{\text{Intra}} + \sigma^{\text{Inter}}$. This can be extracted using the definition of electric current $\bm{j} = -e \text{Tr}[\bm{v} \rho]=\sigma \bm {E}$.  In the present work, we are dealing with the band basis representation and DNLSMs, thus it is better to define velocity for DNLSMs as $ \space \tilde{v}^{pn}_i=\space\delta_{pn}\space\partial_{\tilde{k}_i}\tilde{\varepsilon}_{\bm{{k}}}^n+i\space\mathcal{\tilde{R}}^{pn}_{{\bm k}_i}\space\tilde{\omega}^{pn}$, where $\tilde{v}_i=\hbar  {k}_0 v_i/\varepsilon_0 $ and $\tilde{\omega}^ {pn}= \tilde{\varepsilon}_{{\bm k}}^{p}-\tilde{\varepsilon}_{{\bm k}}^{n}$ is the energy difference between the $p$ and $n$ bands. Here, the first term of the Bloch velocity refers to the intraband contribution or the group velocity and the second term to the interband contribution. Below we explain intraband and interband contributions explicitly.\\
\textit{Intraband Contribution:}
Considering the electric field along $x$-direction, the intraband contribution to the total anomalous conductivity can be obtained following the formula $\sigma_{zx}^{\text{Intra}} E_x= -e \sum_{\bm k} v_{z}^{nn} \mathcal{N}_{\bm E,\bm k}^{nn}$  and using Eq.~\eqref{Eq:N_E}

\begin{align}\label{eqn:intra_ij}
\sigma_{zx}^{\text {Intra} } = -\frac{e^2}{\hbar} \sum_{n}\sum_{{\bm k}} \frac{1}{\tilde{g}_{{\bm k}}+i\tilde{\omega}} \tilde{v}_z^{nn} \tilde{v}_x^{nn}\delta{(\tilde{\varepsilon}_{{\bm k}}^{n}-\tilde{\mu})} ,
\end{align}
where  $\tilde{g}_{\bm k}=\frac{g_{\bm k}}{\varepsilon_0}$, $\tilde{\omega} = \hbar \omega/\varepsilon_0$, and  we can replace $\frac{\partial f^0}{\partial \tilde{k}_z} = \frac{\partial f^{0}}{\partial \tilde{\varepsilon}_{{ k}}}\frac{\partial \tilde{\varepsilon}_{{\bm k}} }{\partial {\tilde{ k}_z}} = - \hbar \tilde{v}_z^{nn} \delta(\tilde{\varepsilon}_{{\bm k}}^n-\tilde{\mu})$ in the low-temperature limit ($T\rightarrow 0$).  
Here we approximate the energy derivative of the Fermi-Dirac distribution function approaches as the Dirac delta function $\delta(\tilde{\varepsilon}_{{\bm {k}}}^n -\tilde{\mu})$ within the low-temperature limit ($T\rightarrow 0$). For the DNLSM, the component of the velocity along $z$-direction yields linear dependence on $k_z$. This makes the whole integrand an odd function in $k_z$, resulting in the zero intraband contribution to the anomalous Hall conductivity.\\
\begin{figure*}[htp]
   \centering
    \includegraphics[width=15 cm]{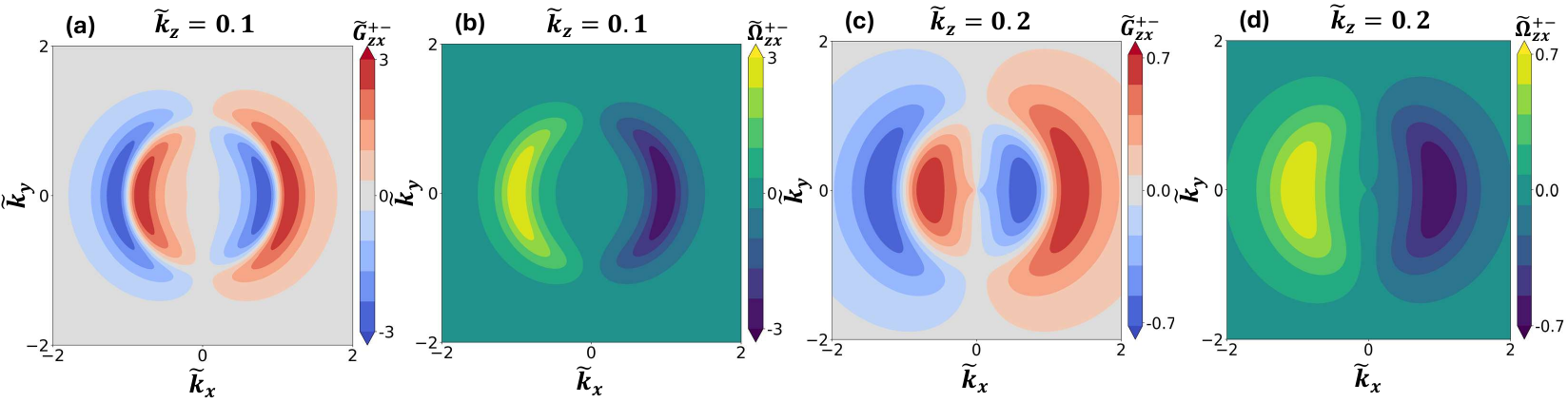}
    \caption{The variation of the quantum metric $\tilde{G}_{zx}^{+-}=Re[\mathcal{\tilde{R}}^{-+}_{k_z}\mathcal{\tilde{R}}^{+-}_{k_x}]$ and Berry curvature $\tilde{\Omega}_{zx}^{+-}=Im[\mathcal{\tilde{R}}^{-+}_{k_z}\mathcal{\tilde{R}}^{+-}_{k_x}]$ for the tilted DNLSM as a function of $\tilde{k}_x$ and $\tilde{k}_y$ at fixed $\tilde{k}_z$ values (a) and (b) $\tilde{k}_z=0.1$; (c) and (d) $\tilde{k}_z = 0.2$.}
    \label{QTG}
\end{figure*}
\textit{Interband Contribution:} The interband part of the conductivity comes from the intrinsic contribution, which is the field-driven contribution, and the extrinsic contribution which stems from the disorder effect (scattering effects) as shown in Eq.~\eqref{eqn:S_E}.\\
First, the extrinsic part of the interband conductivity gives 
\begin{align}
\label{eqn:ext_ij}
\sigma_{zx}^{\text{Ext}} = - \frac{ie}{E_x} \sum_{n\neq p}\sum_{{\bm k}}\tilde{\mathcal{R}}_{{ k}_z}^{np}\frac{\mathcal{J}_{\tilde{{k}}_x}^{np}[\mathcal{N}_{\bm E, \bm k}]\tilde{\omega}^{pn}}{\tilde{g}+\space i\space (\tilde{\omega}^{pn}+ \tilde{\omega})}.
\end{align}
Here, $\tilde{g}=\frac{g}{\varepsilon_0}$, the energy difference factor $\tilde{\omega}^{np} = \tilde{\varepsilon}^{n}_{\bm k}-\tilde{\varepsilon}^{p}_{\bm k}$ dictates the resonance on approaching the difference to the energy $\tilde{\omega}$. Further, we treat the scattering term within the weak disorder limit of $\mu \tau/\hbar \gg 1$ or $k_F l \gg 1$ and consider it up to the first-order Born approximation as expressed in Eq.~\eqref{eqn:ext_ij}.  The extrinsic interband part of anomalous conductivity relies on the partial derivative of the Fermi distribution function with respect to the wave vector, thus termed as the Fermi surface contribution. The latter dependence is embedded within the scattering term $\mathcal{J}_{\bm k}^{np}[\mathcal{N}_{\bm E, \bm k}]$ in Eq.~\eqref{eqn:ext_ij}, where the field correction diagonal part of the density matrix follows $\mathcal{N}_{\bm E, \bm k}\propto \partial f^0_{\bm k}/\partial \bm k= \frac{\partial f^0_{\bm{k}}}{\partial \varepsilon^{n}_{\bm{k}}} \frac{\partial \varepsilon^{n}_{\bm{k}}}{\partial \bm{k}}$.  Moreover, this term incorporates the asymmetry induced by the tilt via the group velocity present in $\mathcal{N}_{\bm E, \bm k}$. For the case of the tilted model system (tilt considered here in $x$ direction), the energy derivative with respect to $k_x$ becomes $\frac{\partial \varepsilon^{n}_{\bm{k}}}{\partial \bm{k}_x} = \tilde{v}_t + n \frac{\tilde{k}_x (\tilde{\mathcal{K}} - 1)}{\tilde{\mathcal{K}} \varepsilon_{\bm{k}}}$, treated the Fermi Dirac distribution function within the low temperature regime $ \frac{\partial f^0_{\bm{k}}}{\partial \varepsilon^{n}_{\bm{k}}} \approx -\delta(\varepsilon^n_{\bm k}-\mu)$ and expand the delta functions for small tilt. One can get nonzero extrinsic AHC only for finite tilt. The more detailed derivation about Eq.~\eqref{eqn:ext_ij} in~\ref{Appendix:A2}. 

Second, the intrinsic interband part of the anomalous conductivity can be written as, 
\begin{align}\label{eqn:int_zx}
\sigma_{zx}^{\text{Int}} =\frac{e^2}{\hbar} \sum_{n\neq p}\sum_{{\bm k}}  \frac{\mathcal{\tilde{Q}}_{zx}^{np} \tilde{F}^{np} \tilde{\omega}^{pn}}{\tilde{g}+\space i \space ( \tilde{\omega}^{pn} + \tilde{\omega})},
\end{align}
where $\tilde{F}^{np}=f^0{(\tilde{\varepsilon}_{\bm{k}}^n)}-f^0{(\tilde{\varepsilon}_{\bm{k}}^p)}$ refers to the occupation probability difference between two separate bands that determines the allowed interband transitions following the Pauli exclusion principle.
Here, $\mathcal{\tilde{Q}}_{zx}^{np}= \mathcal{\tilde{R}}^{pn}_{{k}_{z}}\mathcal{\tilde{R}}^{np}_{{k}_{x}} = \tilde{G}^{np}_{zx} + i/2 \tilde{\Omega}^{np}_{zx}$ is the band-resolved quantum geometric tensor (QGT). Here, the real part of QGT represents the quantum metric ($ \tilde{G}^{np}_{zx}$) which quantifies the change in the wave function under the small change in the wave vector. The imaginary part of QGT refers to the Berry curvature ($\tilde{\Omega}^{np}_{zx}$) that measures the geometrical phase structure in the $\bm k$-space.
For our two-band model Hamiltonian for DNLSM Eq.~\eqref{eqn:hamitonian}, the quantum geometric tensor becomes, $\tilde{\mathcal{Q}}_{zx}^{-+} = \frac{\gamma \tilde{k}_x}{4 \tilde{k} \tilde{\varepsilon}_{\bm k}^3} \{\frac{\gamma \tilde{k}_z (\tilde{\mathcal{K}}-1)}{\tilde{\varepsilon}_{{\bm k}}}-i \tilde{M}\}$, which results the quantum metric $\tilde{G}^{-+}_{zx}= \tilde{G}^{+-}_{zx} =\frac{\gamma^2 \tilde{k}_x \tilde{k}_z (\tilde{\mathcal{K}}-1)}{4 \tilde{\mathcal{K}} \tilde{\varepsilon}_{{\bm k}}^4}$ and the Berry curvature $\tilde{\Omega}^{-+}_{zx}=-\tilde{\Omega}^{+-}_{zx}=\frac{\gamma \tilde{k}_x \tilde{M}}{2 \tilde{\mathcal{K}} \tilde{\varepsilon}_{{\bm k}}^3}$.
It is important to note that both quantum metric and Berry curvature are tilt ($\tilde{v}_t$) independent.
Additionally, for the gapless system $\tilde{M}=0$, the Berry curvature vanishes, while the quantum metric remains non-zero even on turning off the mass term. Further, at $\tilde{k}_z=0$ quantum metric vanishes while the Berry curvature remains finite. The variation of the Berry curvature $\Omega_{zx}^{-+}$ and quantum metric in $k$-space for different $\tilde{k}_z$ values are displayed in Fig.~\ref{QTG}.
In the intrinsic part of conductivity, the tilt dependence originates only from the occupation probability. Here, at the low temperature the factor $\tilde{F}^{np}$ reduces to the difference between the step function. This results in the integration limit for the angular part as $\phi_{\pm}= \cos^{-1}\bigg[\frac{(\tilde{\mu} \pm \tilde{\varepsilon}_{{\bm k}})}{\tilde{v}_t \mathcal{\tilde{K}}}\bigg]$. Moreover, the intrinsic interband part of anomalous conductivity comes from the Fermi sea contribution due to the dependence on the occupancy difference between different bands ($\sigma_{zx}^{\text{Int}}\propto \tilde{F}^{np}$) in Eq.~\eqref{eqn:int_zx} and more details are given in~\ref{Appendix:A3}.

\begin{table*}[htp]
\centering
\caption{Transformation behavior of relevant quantities for anomalous Hall conductivity (AHC) of three dimensional Dirac Nodal line semimetals under inversion symmetry ($\mathcal{P}$) with tilt zero ($v_t=0$) and broken inversion symmetry with finite tilt ($v_t\neq0$)  along the $x$ direction, here $\tilde{\varepsilon}_{t,\bm k}=\tilde{v}_t \tilde{k}_x$ and $\tilde{v}_{t,\bm k}=\tilde{v}_t$. }
\begin{tabular}{|>{\centering\arraybackslash}p{5.2cm}|>{\centering\arraybackslash}p{4.9cm}|>{\centering\arraybackslash}p{4.9cm}|}
\hline
\textbf{Quantity} & \textbf{$\bm{v_t =0}$, $\mathcal{P}$ symmetric}& \textbf{$\bm{v_t \neq 0}$, $\mathcal{P}$ broken}\\
\hline
Band dispersion: $\tilde{\varepsilon}^n_{\bm k}$& $\tilde{\varepsilon}^n_{\bm k} =\tilde{\varepsilon}^n_{-\bm k}$ & $\tilde{\varepsilon}^{n}_{\bm k} = \tilde{\varepsilon}_{t,\bm k}  +n\tilde{\varepsilon}_{\bm k}$, $\tilde{\varepsilon}_{\bm k} =  \tilde{\varepsilon}_{-\bm k}$, $\tilde{\varepsilon}_{t,\bm k} =  -\tilde{\varepsilon}_{t,-\bm k}$\\
\hline
Group velocity: $\tilde{v}^n_{\bm k}$&  $\tilde{v}^n_{\bm k} = -\tilde{v}^n_{\bm k}$ & $\tilde{v}^n_{\bm k} = \tilde{v}_{t, {\bm k}} + n \tilde{v}_{\bm k}$,  $ \tilde{v}_{t, {\bm k}}=  \tilde{v}_{t, {\bm k}}$, $ \tilde{v}_{\bm k}= -  \tilde{v}_{\bm k}$\\
\hline
Intraband density matrix: $\mathcal{N}_{\bm E, \bm k}^{nn}\propto \bm E \cdot \frac{\partial f^0_{\bm k}}{\partial  \bm k}$&  $\mathcal{N}_{\bm E, \bm k}^{nn} = \mathcal{N}_{-\bm E, -\bm k}^{nn}$& $\mathcal{N}_{\bm E, \bm k}^{nn} \neq\mathcal{N}_{-\bm E, -\bm k}^{nn}$ \\
\hline
Scattering term: $\mathcal{J}^{np}_{\bm k}\propto \mathcal{N}_{\bm E, \bm k}^{nn}$& $\mathcal{J}^{np}_{\bm k} = \mathcal{J}^{np}_{-\bm k}$ (even)& $\mathcal{J}^{np}_{\bm k} \neq \mathcal{J}^{np}_{-\bm k}$\\ 
\hline 
 Berry Connection: $\tilde{\mathcal{R}}^{pn}_{k_i}$& $\tilde{\mathcal{R}}^{pn}_{{\bm k}_i} = -\tilde{\mathcal{R}}^{pn}_{-{\bm k}_i}$&$\tilde{\mathcal{R}}^{pn}_{{\bm k}_i} = -\tilde{\mathcal{R}}^{pn}_{-{\bm k}_i}$ \\
\hline
Berry curvature: $\tilde{\Omega}^{np}_{zx}(\bm k)$& $\tilde{\Omega}^{np}_{zx}(\bm k) =\tilde{\Omega}^{np}_{zx}(-\bm k)$& $\tilde{\Omega}^{np}_{zx}(\bm k) = \tilde{\Omega}^{np}_{zx}(-\bm k)$\\
\hline
Quantum metric: $\tilde{G}^{np}_{zx}(\bm k)$& $\tilde{G}^{np}_{zx}(\bm k)$=$\tilde{G}^{np}_{zx}(-\bm k)$& $\tilde{G}^{np}_{zx}(\bm k)$=$\tilde{G}^{np}_{zx}(-\bm k)$ \\ \hline 
 Intraband AHC: $\sigma_{zx}^{\text{Intra}} = -\frac {e}{E_x} \sum_{\bm k} v_{z}^{nn} \mathcal{N}_{\bm E,\bm k}^{nn}$& $\sigma_{zx}^{\text{Intra}} =0$& $\sigma_{zx}^{\text{Intra}} = 0$\\ \hline 
 Intrinsic interband AHC: $\sigma_{zx}^{\text{Int}} =\frac{e^2}{\hbar} \sum_{n\neq p}\sum_{{\bm k}}  \frac{\mathcal{\tilde{Q}}_{zx}^{np} \tilde{F}^{np} \tilde{\omega}^{pn}}{\tilde{g}+\space i \space ( \tilde{\omega}^{pn} + \tilde{\omega})}$& $\sigma_{zx}^{\text{Int}} \neq 0$ ($\mathcal{T}$ broken at each point in the nodal ring)  \par                         $\sigma_{zx}^{\text{Int}} = 0$ ($\mathcal{T}$ broken over whole nodal ring)& $\sigma_{zx}^{\text{Int}} \neq 0$ ($\mathcal{T}$ broken at each point in the nodal ring) \par                              $\sigma_{zx}^{\text{Int}} \neq 0$ ($\mathcal{T}$ broken over whole nodal ring)\\ \hline 
 Extrinsic interband AHC: $\sigma_{zx}^{\text{Ext}} = - \frac{ie}{E_x} \sum_{n\neq p}\sum_{{\bm k}}\tilde{\mathcal{R}}_{{ k}_z}^{np}\frac{\mathcal{J}_{\tilde{{k}}_x}^{np}[\mathcal{N}_{\bm E, \bm k}]\tilde{\omega}^{pn}}{\tilde{g}+\space i\space (\tilde{\omega}^{pn}+ \tilde{\omega})}$& $\sigma_{zx}^{\text{Ext}} \neq 0$ ($\mathcal{T}$ broken at each point in the nodal ring)                               $\sigma_{zx}^{\text{Ext}}  = 0$ ($\mathcal{T}$ broken over whole nodal ring)& $\sigma_{zx}^{\text{Ext}} \neq 0$ ($\mathcal{T}$ broken at each point in the nodal ring)                               $\sigma_{zx}^{\text{Ext}}  \neq 0$ ($\mathcal{T}$ broken over whole nodal ring)\\ \hline
\end{tabular}
\label{tab:inversion_symmetry}
\end{table*}

To get more physical insights about the extrinsic and intrinsic contributions we can consider the DC case ($\omega \rightarrow 0$) and clean limit ($1/\tau \rightarrow 0$). In this case, the extrinsic interband part of the anomalous conductivity yields $\sigma_{zx}^{\text{Ext}}\propto \tilde{\mathcal{R}}_{{k}_z}^{np} \mathcal{J}_{\tilde{k}_x}^{np}[\mathcal{N}_{\bm E, \bm k}]$. This shows a dependence of interband contribution on the intraband part of the density matrix.
Similarly,  the intrinsic interband part of the anomalous conductivity adheres to $\sigma_{zx}^{\text{Int}}\propto \tilde{\mathcal{Q}}_{zx}^{np} \tilde{F}^{np}$. This shows the direct connection of the intrinsic part with the topological aspects of DNLSMs via the quantum geometric tensor. Thus both the extrinsic and intrinsic terms reflect the interplay of scattering (extrinsic) and geometrical (intrinsic) effects which are important for the comprehensive understanding of the total electronic behavior of the DNLSMs.

\subsection{Inversion symmetry analysis} \label{subsec:symmetry}
In this subsection, we discuss the symmetry analysis of the intraband and interband parts of the conductivity.
First, under the inversion symmetry ($\mathcal{P}$), the band dispersion of DNLSM (absence of tilt) remains unchanged on replacing $\tilde{\bm{k}} \rightarrow -\tilde{\bm{k}}$ and follows $\tilde{\varepsilon}^n_{{\bm{k}}} \rightarrow \tilde{\varepsilon}^n_{-{\bm{k}}}$. However, the velocity reverses the sign under inversion symmetry, hence the product of two velocities $\tilde{v}_{z} \tilde{v}_{x}$, remains invariant under inversion symmetry. 

On breaking the inversion symmetry of the system by adding the tilt term, the band dispersion shows $   \tilde{\varepsilon}^n_{{\bm k}} = \tilde{v}_t \tilde{k}_x + n \tilde{\varepsilon}_{{\bm k}}$. Here the first term flips the sign while the second term remains the same under inversion symmetry. Group velocity $\tilde{v}^n_{\bm k}$ also possesses two terms as  $\tilde{v}^n_{\bm k} = \tilde{v}_{t, \bm k} + n\tilde{\bm k}/\tilde{\varepsilon}_{\bm k}$, where the first term preserves the sign while the second changes the sign under $\mathcal{P}$ operation it is important to note that tilt is considered along  $x$ direction. However the other components of velocity such as $\tilde{v}_y$ and $\tilde{v}_z$ flip the sign. In this case, the tilt dependent factor of  $\tilde{v}_z\tilde{v}_x$ shows sign change under inversion symmetry operation. 
This results in the nonzero intraband part of the anomalous conductivity.  However in the present case for tilted DNLSM, the intraband part vanishes due to the linear wave vector dependence along $z$-direction as stated earlier.

Second, in the case of the extrinsic interband part of the anomalous conductivity, the intraband part of the density matrix preserves the sign as $\mathcal{N}_{\bm E, \bm k}^{nn}= \mathcal{N}_{-\bm E, -\bm k}^{nn}$, hence the scattering term follows $\mathcal{J}^{np}_{\bm{k}} = \mathcal{J}^{np}_{\bm{-k}}$. This results zero anomalous Hall conductivity.
However, in the presence of tilt the intraband part of the density matrix flips the sign $\mathcal{N}_{\bm E, \bm k}^{nn}\neq\mathcal{N}_{-\bm E, -\bm k}^{nn} $, which can cause $\mathcal{J}^{np}_{\bm{k}} \neq\mathcal{J}^{np}_{\bm{-k}}$. Hence we will get a non-zero extrinsic interband part of the anomalous Hall response($\tilde{\sigma}^{\text{Ext}}_{zx}\neq 0$).

Third for the intrinsic interband conductivity, the expression contains the quantum geometric tensor (QGT), defined as $\tilde{\mathcal{Q}}_{zx}^{np} = \tilde{G}_{zx}^{np} + \frac{i}{2}\tilde{\Omega}_{zx}^{np}$. Under inversion symmetry, the Berry connection $\tilde{\mathcal{R}}^{pn}_{k_i}$ changes sign as $\tilde{\mathcal{R}}^{pn}_{{\bm k}_i} = -\tilde{\mathcal{R}}^{pn}_{-{\bm k}_i}$, while both the geometric quantities such as Berry curvature $\tilde{\Omega}_{zx}^{np}(\bm{k}) = \tilde{\Omega}_{zx}^{np}(\bm{-k})$ and quantum metric $\tilde{G}_{zx}^{np}(\bm{k}) = \tilde{G}_{zx}^{np}(\bm{-k})$ remain unchanged, whereas, $\int d^3 \bm k \rightarrow -\int d^3 \bm k $ changes its sign. Therefore, the intrinsic conductivity $\tilde{\sigma}^{\text{Int}}_{zx}= 0$ for an inversion symmetric system, resulting zero anomalous Hall conductivity. Thus the non zero intrinsic part of the anomalous Hall conductivity requires a broken inversion symmetry system. Here, the tilt-dependent term (an inversion symmetry broken term) arises from the band dispersion dependent Fermi distribution function making the intrinsic interband part of the anomalous conductivity nonzero. The details about the symmetry analysis are given in Table~\ref{tab:inversion_symmetry}.
\begin{figure*}[htp]
    \centering
    \includegraphics[width=15cm]{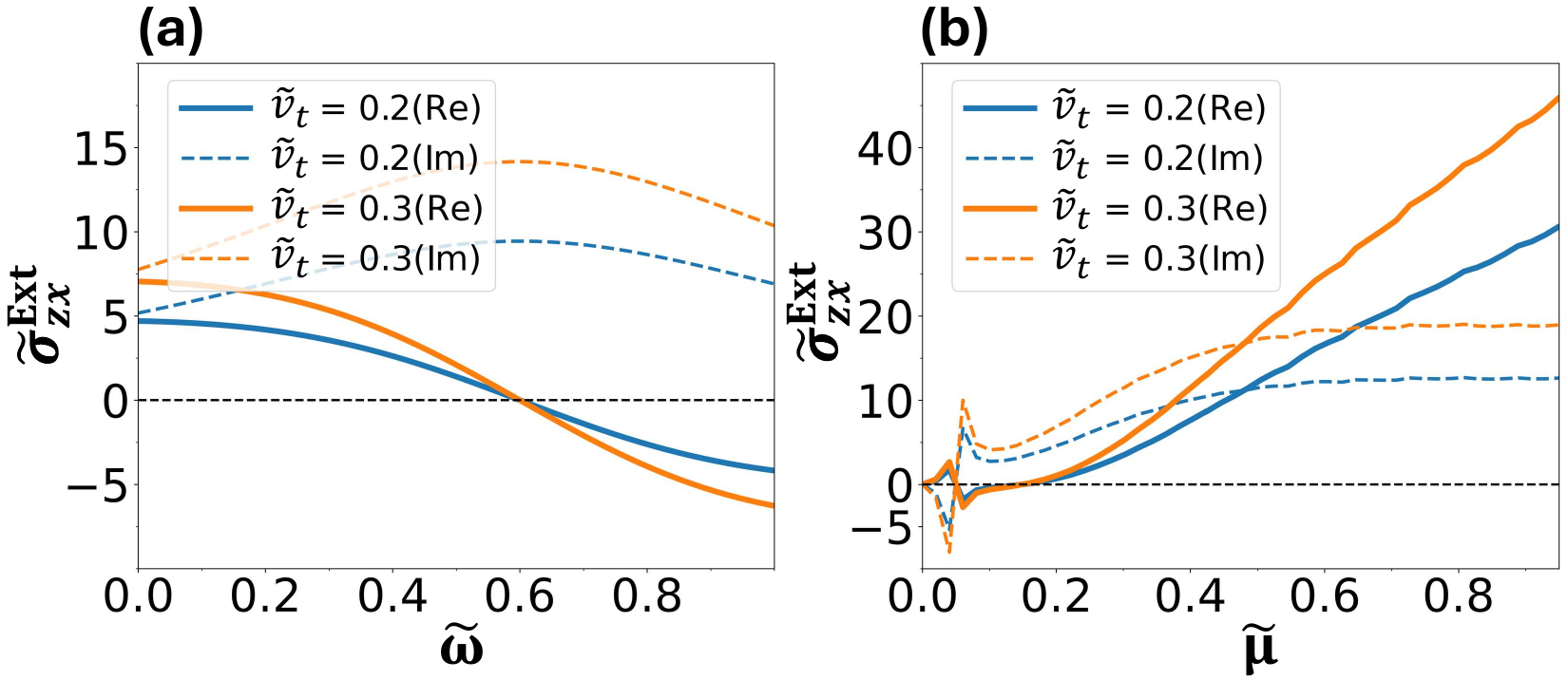}
    \caption{(a) shows the real part (solid line) and imaginary part (dotted line) of the extrinsic interband part of anomalous conductivity $\tilde{\sigma}_{zx}^{\text{Ext}}=\sigma_{zx}^{\text{Ext}}/\sigma_0$ where $\sigma_0 = e^2 k_0/\hbar(2\pi)^2 $ as a function of frequency ($\tilde{\omega}$) at a fixed chemical potential $\tilde{\mu}=0.3$, (b) depicts the extrinsic interband response with the chemical potential ($\tilde{\mu}$) at a fixed frequency $\tilde{\omega}=0.3$. The results are shown for tilt parameters $\tilde{v}_t=0.3$ (orange) and $\tilde{v}_t=0.2$ (blue), respectively.} 
    \label{fig:extrinsic}
\end{figure*}

\section{Results and Discussion}
\label{sec: results}
In the Fig.~\ref{fig:extrinsic}, we show the variation of the interband part of the anomalous conductivity for the inversion symmetry broken DNLSM system with respect to the frequency and chemical potential at distinct tilt values. Here, the solid curves refer to the real part and the dashed curves refer to the imaginary part of the anomalous Hall conductivity. 

We find that the extrinsic interband part of the anomalous conductivity is presented in Fig.~\ref{fig:extrinsic}, illustrating the variation of $\tilde{\sigma}_{zx}^{\text{Ext}}$ with respect to frequency $\tilde{\omega}$ and chemical potential $\tilde{\mu}$ for different values of the tilt $\tilde{v}_t$. In Fig.~\ref{fig:extrinsic}(a), we observe a non-zero response at $\tilde{\omega}=0$ (DC limit) due to the direct dependence of extrinsic interband response on $v_t$. This dependence arises from the term $\mathcal{J}_{\tilde{k}_x}^{np}[\mathcal{N}_{\bm E, \bm k}]$ in Eq.~\eqref{eqn:ext_ij}, which is directly proportional to the term governing the gradient of the equilibrium distribution function in $\bm k$-space ($ \partial f^0_{\tilde{\bm k}}/\partial\tilde{\bm k}$). In addition, the transition between the electron to hole plateau leads to the sign conversion in the extrinsic intraband response at $\tilde{\omega}=2\tilde{\mu}$. Further, an increase in $\tilde{v}_t$ enhances the magnitude of the real part of the extrinsic interband response, while the overall behavior of the system remains intact. Corresponding to the real part, the imaginary part of the extrinsic interband part of the anomalous response $Im[\tilde{\sigma}_{zx}^{\text{Ext}}]$ shows a hump at a frequency equal to twice the value of the chemical potential. The given discussion is consistent with Eq.~\eqref{eqn:ext_ij}, where  $\tilde{\sigma}_{zx}^{\text{Ext}}$ is directly proportional to tilt $v_t$ dependent scattering term at $\tilde{\omega} \rightarrow 0$ and $\tilde{g} \rightarrow 0$, which reveals the importance of taking the disorder effect in the system to understand the net response of the system. Note that here $\tilde{\sigma}_{zx}^{\text{Ext}}=0$ at $\tilde{v}_t=0$. At finite $\tilde{\omega}$, mathematically, the behavior with frequency is described by the factor $\frac{1}{\tilde{g}+\space i\space (\tilde{\omega}^{pn}+ \tilde{\omega})}$ in Eq.~\eqref{eqn:ext_ij}. Notably, here the kinks at the characteristic frequencies analogous to the intrinsic part do not occur due to the origin of the extrinsic component of the interband part of conductivity from the Fermi surface effect. 

To illustrate the chemical potential dependence, we show a plot $\tilde{\sigma}_{zx}^{\text{Ext}}$ versus $\tilde{\mu}$ at different $\tilde{v}_t$ in Fig.~\ref{fig:extrinsic}(b). We find small variations in the response at low chemical potential values when the chemical potential touches the bottom of the conduction band. However, at $\tilde{\mu} \gg \tilde{M}$, the extrinsic interband part of conductivity grows linearly. Furthermore, the tilt parameter ($\tilde{v}_t$) influences the magnitude of the response, causing it to increase as $\tilde{v}_t$ becomes larger. On the other hand, the overall qualitative behavior of the response remains intact. 
\begin{figure*}[htp]
    \centering
    \includegraphics[width=15 cm]{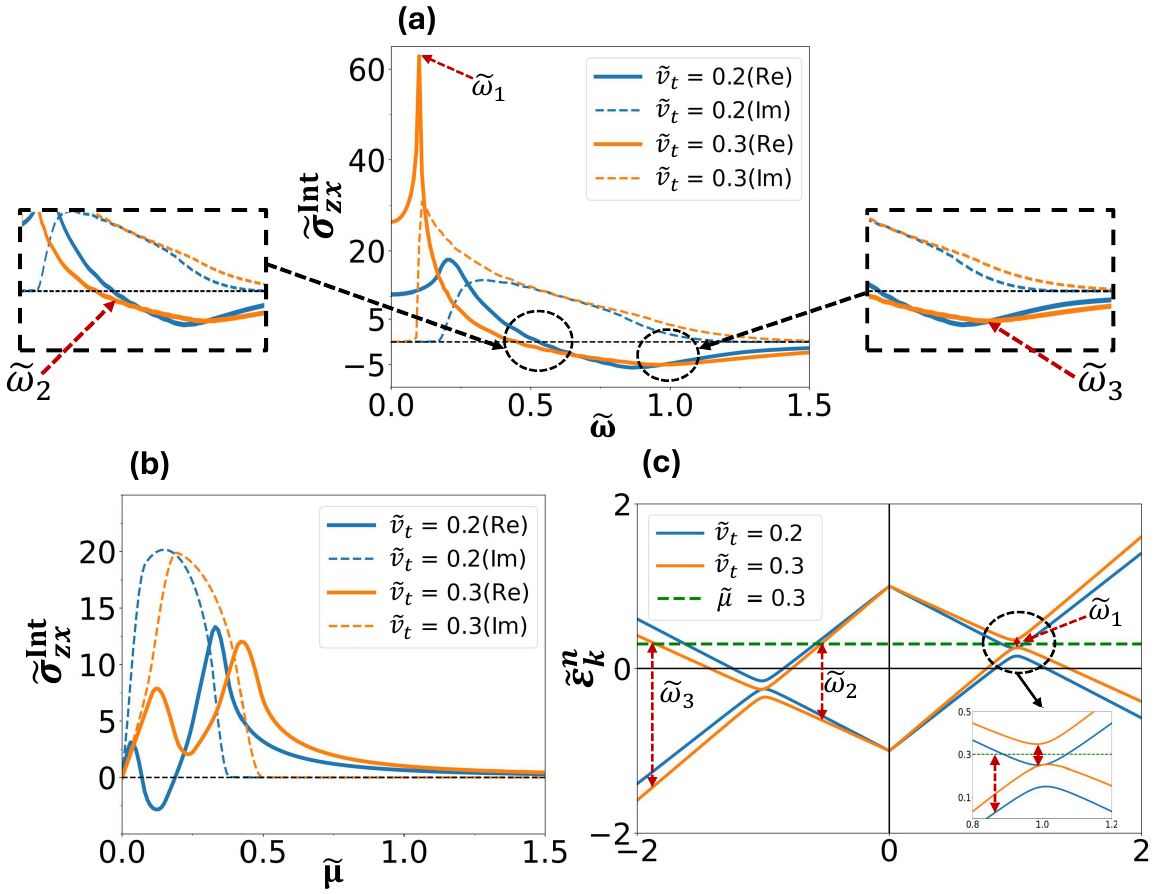}
    \caption{(a) illustrates the real part (solid line) and the imaginary part (dotted line) of the intrinsic interband part of anomalous conductivity $\tilde{\sigma}_{zx}^{\text{Int}}=\sigma_{zx}^{\text{Int}}/\sigma_0$ where $\sigma_0 = e^2 k_0/\hbar(2\pi)^2 $ as a function of frequency ($\tilde{\omega}$) at $\tilde{\mu}=0.3$. Here, the red arrows correspond to the kinks that are associated with the characteristic frequencies $\tilde{\omega}_1$, $\tilde{\omega}_2$ and $\tilde{\omega}_3$ for tilt $\tilde{v}_t = 0.3$. (b) refers to the intrinsic interband part of anomalous conductivity as a function of the chemical potential ($\tilde{\mu}$) at $\tilde{\omega}=0.3$. (c) represents the one-dimensional dispersion plot to explain the kinks observed in (a). Further here the chemical potential is set as $\tilde{\mu}=0.3$ characteristic frequencies $\tilde{\omega}_1$, $\tilde{\omega}_2$ and $\tilde{\omega}_3$. The zoom version of the dispersion within the black circle is shown in the inset of the figure. Similar kinks for the tilt $\tilde{v}_t=0.2$ can also be seen at distinct characteristic frequencies. Notably, the results correspond to tilt parameters $\tilde{v}_t=0.3$ (orange) and $\tilde{v}_t = 0.2$ (blue), respectively.} 
    \label{fig:intrinsic}
\end{figure*}

In the intrinsic interband part of the conductivity depicted in Fig.~\ref{fig:intrinsic} gives a non-zero contribution in the presence of tilt. However, it vanishes in the absence of tilt. This happens due to the broken inversion symmetry by the tilt contribution.  
Furthermore, the tilt dependence of $\tilde{\sigma}_{zx}^{\text{Int}}$ comes entirely from the occupancy factor $\tilde{F}^{np}= f(\tilde{\varepsilon}_{{\bm k}}^n)- f(\tilde{\varepsilon}_{{\bm k}}^p)$ where the tilt is incorporated within the dispersion. In addition, the real part of the intrinsic interband contribution in the DC limit gives a finite value as shown in Fig.~\ref{fig:intrinsic}(a).
Further, we observe that the intrinsic interband part of the conductivity (Fig.~\ref{fig:intrinsic}(a)) follows a peculiar behavior as a function of $\tilde{\omega}$ by showing kinks at different frequencies. This is related to the transitions at different $\tilde{\omega}$ values depending on the competition between the chemical potential and the tilt value. Here we find three kinks associated with frequencies $\tilde{\omega}_1$,  $\tilde{\omega}_2$ and  $\tilde{\omega}_3$. The values of these characteristic frequencies can be qualitatively described via Fig.~\ref{fig:intrinsic}(c) where we have shown the dispersion as a function of $\tilde{k}_x$ with solid lines and the chemical potential by dashed line. The prominent peak (orange curve) occurs at $\tilde{\omega}_1=2\tilde{M}$, the band gap between the conduction and the valence band as shown in Fig.~\ref{fig:intrinsic}(a). Note that the location of the chemical potential lies in the band gap region. However, the other peaks (small orange kinks) and all blue color peaks occur at later frequencies where the chemical potential lies within the conduction band.
There is also a sign change in the real part of intrinsic interband conductivity as the external energy ($\tilde{\omega}$) becomes larger than the gap $2\tilde{M}$. At higher frequencies, it approaches zero due to inverse variation with $\tilde{\omega}$.
At a higher magnitude of $\tilde{v}_t$, the magnitude of the intrinsic response increases whereas the behavior remains unchanged. It is important to note that the real part of the intrinsic interband part of the anomalous conductivity at high frequency becomes tilt-independent.
Similar scenarios can be observed in the imaginary part of the intrinsic contribution of the anomalous conductivity ($Im[\tilde{\sigma}_{zx}^{\text{Int}}]$ ).

Fig.~\ref{fig:intrinsic}(b) depicts the anomalous response $\tilde{\sigma}_{zx}^{\text{Int}}$ as a function of scaled chemical potential $\tilde{\mu}$ at different tilt values. At $\tilde{\mu}=0$, the intrinsic interband part of the conductivity shows no response due to the zero contribution made by $\tilde{\mu}$ dependent azimuthal angle part. The increase in the doping causes the chemical potential to shift, hence modifying the characteristic frequencies and the respective peaks. The location of the observed peaks depends on the position of the chemical potential as shown in Fig.~\ref{fig:intrinsic}(c). For $\tilde{v}_t = 0.3$ an increase in the chemical potential shifts the chemical potential into the conduction band, causing the location of the peaks to differ from the peaks observed in Fig.~\ref{fig:intrinsic}(a). A similar trend is observed in the case of $\tilde{v}_t = 0.2$, where at low frequencies, the chemical potential lies within the gap region, but at higher frequencies, it enters inside the conduction band, resulting in a corresponding shift in the observed peak location.  
In the case of the imaginary part of the intrinsic interband contribution, $Im[\tilde{\sigma}_{zx}^{\text{Int}}]$ the system response shows a non-monotonic behavior.    
It is important to note that at high frequencies, the intrinsic interband part of the response becomes tilt-independent and moves towards zero.

\subsection{Validity:} \label{subsec:validity}
Our study is valid under the condition $\mu\tau/\hbar \gg 1$, which corresponds to a weak disorder limit. On considering the interband relaxation time $\tau \approx 10^{-12}$s~\cite{Rudenko_prb2020, Schilling_prl2017, Zoltan_prb2021} and $\hbar =  6.6 \times 10^{-16}$ eV, we find $\mu \gg 0.6 \times 10^{-3}$. In the scaled form, our results are valid for $\tilde{\mu} \gg 0.003$ eV. In this regime, the perturbative theory assumes that weak disorder introduces a small modification to the total Hamiltonian, which governs the energy distribution. In the present study, we employed the first-order Born approximation for the treatment of weak disorder effects.
Within this approach, one can keep finite order terms based on the strength of the disorder and is widely used to understand how disorder influences electronic properties, such as conductivity. Specifically, the approach is effective when the disorder is weak to moderate, but as the disorder intensifies, the first-order Born approximation does not remain valid. 
To observe the impact of higher-order potential one can go beyond the first-order Born approximation. In the case of strong disorder $\mu\tau/\hbar \ll 1$, the Bloch band picture breaks down and the Fermi surface as well as the topological features become ill-defined (not sharply defined) and the first-order Born approximation will no longer be valid. 
Hence for the higher-order potential one has to go beyond the first order Born approximation.

Furthermore, our calculation is performed under the low temperature limit plays a role in understanding the sharp features, such as the kinks at the distinct characteristic frequency associated with the interband response. On the other hand, at a high temperature, the Fermi Dirac distribution smooths (broadens), leading to a smooth variation in the occupation of the states near the chemical potential. Thus, transitions that were sharply getting turned on or off at $T\rightarrow 0$ become gradual over the energy range $k_B T$ and it leads to the smearing (broadening) of the frequency dependent kinks. 
In addition, the occupancy difference $f(\varepsilon^n_{\bm k})-f(\varepsilon^p_{\bm k})$ is smoothed at a finite temperature. As a result, the interband transitions can occur in the broad frequency range instead of a sharp threshold $\omega^{np}=\varepsilon^n_{\bm k}-\varepsilon^p_{\bm k}$, which can shift the onset of interband transitions. Additionally, at high temperatures, the electron-phonon scattering becomes prominent over the impurity scattering, which may modify the anomalous Hall response of the system. Such scattering effect can be incorporated by taking the electron-phonon interaction part in the Hamiltonian, which will modify the scattering-driven part of the density matrix $\mathcal{J}_{\bm k}^{np}[\mathcal{N}_{E,\bm k}]$, hence the extrinsic part of the interband conductivity of the system.
However, at low temperature case electron-phonon interaction can be neglected~\cite{Rudenko_prb2020}.

Our study considers the low energy $k\cdot p$ Hamiltonian, which captures the important features of DNLSMs in the vicinity of the nodal ring. This model explains the underlying physics of the system near the Fermi level. To consider the effects of other bands away in the momentum space, one has to go with the ab intio approach which captures the electronic structure over the entire Brillouin zone. This may add additional contributions to AHC via the scattering driven matrix elements $U_{\bm k, \bm {k}'}^{np}= \bra{n,\bm k}U(\bm r)\ket{p,\bm {k}'}$, modified Berry curvature $\Omega^{np}(\bm k)$ and the interband transitions within the remote bands.

\subsection{Experimental Relevance and Numerical estimation:} \label{subsec:Experimental relevance}
To investigate the anomalous Hall conductivity in the tilted DNLSMs, the sample is connected to the voltage source by applying a bias along the $x$-direction and the response along $z$-direction can be measured. Further, by shifting the chemical potential using the gate techniques and electronic doping the resulting response can be modulated and dictated theoretically by the intrinsic and extrinsic contributions.
The intrinsic and extrinsic contributions in real materials such as ZrSiS or Ca$_3$P$_2$ can be distinguished by utilizing frequency-resolved optical conductivity measurements. 
The intrinsic response, governed by the band geometry (e.g., Berry curvature and quantum metric), typically exhibits tunable sharp onset features or kinks associated with interband transitions with the chemical potential and tilt, allowing for direct comparison with theoretical predictions. 
In contrast, the extrinsic response, arising from impurity scattering, is more sensitive to disorder and often appears as a broader background contribution that may not exhibit sharp thresholds (decay or flatten) but can vary with impurity concentration or relaxation time. One can suppress or enhance the extrinsic effects and isolate the intrinsic response by performing disorder-dependent optical conductivity measurements.

For experimental relevance, the numerical estimation of both extrinsic and intrinsic contributions to the interband part of anomalous Hall response for DNLSM Ca$_3$P$_2$~\cite{Xie_APLM2015, Chan_prb2016} can be obtained as follows. Based on density functional theory (DFT) estimated parameters for $\text{Ca}_3\text{P}_2$, we consider the radius of the nodal ring $k_0 \approx 0.206 \AA^{-1}$, the energy associated with the radius of the ring $\varepsilon_0 \approx 0.184$ eV the relaxation time scale $\tau \sim  (10^{-12}- 10^{-14})$s and $\gamma \approx 2.80$. Further, we have taken the tilt parameter $\tilde{v}_t = 0.2$ which leads to the velocity $v_t = 0.2 \times 10^{5}$ms$^{-1}$, which is smaller than the order of Fermi velocity ($\sim 10^5$ms$^{-1}$). Using these values and setting frequency $\tilde{\omega}=0.3$ ($\omega \approx 0.3\varepsilon_0$, where $\varepsilon_0 = 0.184 eV$) and chemical potential $\tilde{\mu}=0.4$ ($\tilde{\mu}=\mu/\varepsilon_0$), both the extrinsic and intrinsic contributions of the total interband response give equal value $\sigma_{zx}^{\text{Int}} = \sigma_{zx}^{\text{Ext}}=11\sigma_0$, where $\sigma_0 = e^2 k_0/(2\pi)^2\hbar$. This gives the total anomalous response as $\sigma_{zx}^{\text{Total}} \approx 22\sigma_0 \approx 27 ~ \Omega^{-1}\text{mm}^{-1}$. On tuning $\tilde{\mu}=0.5$ the extrinsic contribution outweighs the intrinsic part of the anomalous response and approximately follows $\sigma_{zx}^{\text{Ext}} \approx 3 \sigma_{zx}^{\text{Int}} = 18 \sigma_0$ and the total response as $\sigma_{zx}^{\text{Total}} \approx 19\sigma_0$ which comes out to be $\sigma_{zx}^{\text{Total}} \approx 23 ~ \Omega^{-1}\text{mm}^{-1}$.

\section{Conclusion}
\label{sec: conclusion}
In conclusion, we have investigated the anomalous Hall conductivity in tilted DNLSMs in the presence of the external electric field, by applying the quantum kinetic approach. Here the total conductivity comprises intraband and interband parts of the conductivity. The interband part can be further classified into two subparts. One is extrinsic part of the interband conductivity, which comes from the scattering-driven off-diagonal part of the density matrix and represents the Fermi Surface contribution. The second is the intrinsic part of the conductivity, which comes from the off-diagonal part of the density matrix, incorporating the external field effects and representing the Fermi sea contribution. We find that both contributions of interband conductivity have significant effects on the total response of the system. Further, observations show that the extrinsic part of the conductivity increases linearly with the chemical potential. On the other hand, the intrinsic conductivity yields different kinks at distinct characteristic frequencies due to the effect of tilt, and gives a peak at $2\tilde{M}$. In addition, the magnitude of both extrinsic and intrinsic contributions increases with the tilt. Furthermore, we find the extrinsic contribution outweighs the intrinsic part, thus making the disorder contribution important to understand the overall signal of the system. Moreover, our results can be tested experimentally, which will be useful in the development of future devices.

\section*{Acknowledgment}
This work is financially supported by the Science and Engineering Research Board-State University Research Excellence under project number SUR/2022/000289.

\onecolumngrid
\appendix
\section{Derivation of Equations} \label{Appendix:A}
\subsection{Derivation of Eq.(9)}
\label{Appendix:A1}
Following Eq.~\eqref{eqn:6} for $S_{\bm E,{\bm k}}^{np}$ in the revised manuscript, we have
\begin{equation} \label{eqn:A6}
    \frac{\partial  S_{\bm E,{\bm{k}}}^{np}}{\partial t}+\frac{i}{\hbar} [ \mathcal{H}_0, S_{\bm E,{\bm{k}}}]^{np}+\mathcal{J}_{\bm{k}}^{np}[\mathcal{N}_{\bm E,{\bm k}}] + \frac{S_{\bm E,{\bm{k}}}^{np}}{\tau}={D_{\bm E,{\bm{k}}}^{np}},
\end{equation}
where the commutator between the band Hamiltonian $\mathcal{H}_0$ and $S_{\bm E, \bm k}$ within the band basis representation gives $\langle n| [ \mathcal{H}_0, S_{\bm E,{\bm{k}}}]|p\rangle= S^{np}_{\bm E,{\bm{k}}}(\varepsilon_{\bm k}^{n}-\varepsilon_{\bm k}^{p}) $, $D_{\bm E,{\bm{k}}}^{np}$ is the field driving term and the scattering term $\mathcal{J}_{\bm{k}}^{np}[\rho_{\bm E,{\bm k}}] = \mathcal{J}_{\bm{k}}^{np}[\mathcal{N}_{\bm E,{\bm k}}]+\mathcal{J}_{\bm{k}}^{np}[\mathcal{S}_{\bm E,{\bm k}}]$ where the first part takes into account the scattering effects via the intraband contributions which we have treated within the first order Born approximation in the later calculations and the second part to the scattering effects via interband contributions which we have considered within the relaxation time approximation as $ \mathcal{S}_{\bm E,{\bm{k}}}^{np}/\tau$  having $\tau$ represents the relaxation time associated with the interband effects and is treated as a constant. 
With this, we obtain
\begin{equation} 
    \frac{\partial  S_{\bm E,{\bm{k}}}^{np}}{\partial t}+S^{np}_{\bm E,{\bm{k}}}\bigg[\frac{i}{\hbar} (\varepsilon_{\bm k}^{n}-\varepsilon_{\bm k}^{p})  + \frac{1}{\tau}\bigg]={D_{\bm E ,{\bm{k}}}^{np}}-\mathcal{J}_{\bm{k}}^{np}[\mathcal{N}_{\bm E,{\bm k}}].
\end{equation}
This is a first order differential equation whose solution becomes
\begin{align}\nonumber
S_{\bm E,\bm{k}}^{np}=\frac{1}{\hbar^2}\int_0^\infty dt' e^{-i\mathcal(\varepsilon_{\bm k}^{n}-\varepsilon_{\bm k}^{p}) t'/\hbar} \space \big({D_{\bm E,{\bm{k}}}^{np}} -\mathcal{J}_{\bm{k}}^{np}[\mathcal{N}_{\bm E,{\bm k}}]  \big)e^{i\mathcal(\varepsilon_{\bm k}^{n}-\varepsilon_{\bm k}^{p}) t'/\hbar}.
\end{align} 
Further, on extracting the time dependent part of the external electric field $\bm E (t)= \bm E e^{-i \omega t}$ stemming from the field driven and scattering driven terms, the integration over time yields
\begin{align}\nonumber
S_{\bm E,\bm{k}}^{np}=\frac{{D_{\bm E,{\bm{k}}}^{np}} -\mathcal{J}_{\bm{k}}^{np}[\mathcal{N}_{\bm E,{\bm k}}] }{\frac{\hbar}{\tau}+i \big(\varepsilon_{\bm{k}'}^n-\varepsilon_{\bm{k}}^p+ \hbar \omega\big)}.
\end{align} 
In compact form, it can be expressed as  
\begin{align}\nonumber
S_{\bm E,\bm{k}}^{np}=\frac{{D_{\bm E,{\bm{k}}}^{np}} -\mathcal{J}_{\bm{k}}^{np}[\mathcal{N}_{\bm E,{\bm k}}] }{g+i \big(\varepsilon_{\bm{k}'}^n-\varepsilon_{\bm{k}}^p+ \hbar \omega\big)}=\frac{{D_{\bm E,{\bm{k}}}^{np}} }{g+i \big(\varepsilon_{\bm{k}'}^n-\varepsilon_{\bm{k}}^p+ \hbar \omega\big)}-\frac{\mathcal{J}_{\bm{k}}^{np}[\mathcal{N}_{\bm E,{\bm k}}] }{g+i \big(\varepsilon_{\bm{k}'}^n-\varepsilon_{\bm{k}}^p+ \hbar \omega\big)}.
\end{align} 
This derives the Eq.~\eqref{eqn:S_E} of the revised manuscript where the first part refers to the intrinsic or field driven and and the second part to the extrinsic or scattering driven.
\subsection{Derivation of Eq.(11) }\label{Appendix:A2}
Similarly, the extrinsic part of the interband conductivity in terms of normalized factor can be written as 
\begin{align}
\label{eqn:Aext_ij}
\sigma_{zx}^{\text{Ext}} =- \frac{ie}{E_x} \sum_{n\neq p}\sum_{{\bm k}}\tilde{\mathcal{R}}_{{ k}_z}^{pn}\frac{\mathcal{J}_{\tilde{{k}}_x}^{np}[\mathcal{N}_{\bm E, \bm k}]\tilde{\omega}^{pn}}{\tilde{g}+\space i\space (\tilde{\omega}^{pn}+ \tilde{\omega})}.
\end{align}
The scattering term in the weak disorder limit on band basis representation is defined in the form by employing the first order Born approximation,
\begin{align}\nonumber
\mathcal{J}_{\bm{k}}^{np}[\mathcal{N}_{\bm E, \bm k}]=\frac{1}{\hbar^2}\int_0^\infty dt' \sum \big[\langle U_{\bm{kk}'}^{nq}U_{\bm{k}'\bm{k}}^{qn}\rangle\space\mathcal{N}_{\bm E,\bm{k}}^{qp}e^{i\big(\varepsilon_{\bm{k}'}^q-\varepsilon_{\bm{k}}^n\big)t'/\hbar}-\langle U_{\bm{kk}'}^{nq}U_{\bm{k}'\bm{k}}^{np} \rangle\space\mathcal{N}_{\bm E,\bm{k}'}^{qn}e^{i\big(\varepsilon_{\bm{k}'}^n-\varepsilon_{\bm{k}}^p\big)t'/\hbar}\\
-\langle U_{\bm{kk}'}^{nq}U_{\bm{k}'\bm{k}}^{np}\rangle\space\mathcal{N}^{qn}_{\bm E,\bm{k}'}e^{i\big(\varepsilon_{\bm{k}'}^n-\varepsilon_{\bm{k}}^q\big)t'/\hbar}+ \langle U_{\bm{kk}'}^{qn}U_{\bm{k}'\bm{k}}^{np}\rangle\space\mathcal{N}^{nq}_{\bm E,\bm{k}}e^{i\big(\varepsilon_{\bm{k}'}^q-\varepsilon_{\bm{k}}^n\big)t'/\hbar}\big],
\end{align} 
where $\langle U_{\bm{kk}'}^{nq} U_{\bm{k}'\bm{k}}^{qt}\rangle = U_0^2 \langle u_{\bm{k}}^n | u_{\bm{k}'}^q \rangle \langle u_{\bm{k}'}^q | u_{\bm{k}}^t \rangle$. Furthermore, the final expression is obtained by regularizing the time integral by inserting a $e^{-\eta {t}'}$ convergence factor and using 

\begin{align}
\frac{1}{\hbar^2}\int_0^\infty dt'   e^{i\big(\varepsilon_{\bm{k}'}^q-\varepsilon_{\bm{k}}^n\big)t'/\hbar} = \mathcal{P}\bigg[\frac{1}{i \hbar\big(\varepsilon_{\bm{k}'}^q-\varepsilon_{\bm{k}}^n\big)}\bigg]+\frac{\pi}{\hbar} \delta\big(\varepsilon_{\bm{k}'}^q-\varepsilon_{\bm{k}}^n\big)
\end{align} 
Here, the principle part accounts for the disorder induced repulsion and the delta function part enforces energy conservation for real scattering events.
Here by considering the scattering events coming from the delta function part, we obtain
\begin{align}\nonumber
\mathcal{J}_{\bm{k}}^{np}[\mathcal{N}_{\bm E, \bm k}]
= \frac{\pi n_i}{\hbar} \sum_{q,\bm{k}'} \bigg\{ 
\langle U_{\bm{kk}'}^{nq}U_{\bm{k}'\bm{k}}^{qn}\rangle\space\mathcal{N}_{\bm E,\bm{k}}^{qp}\space\delta(\varepsilon_{\bm{k}'}^{q} - \varepsilon_{\bm{k}}^{n}) \notag 
 +\langle U_{\bm{kk}'}^{nq}U_{\bm{k}'\bm{k}}^{np}\rangle\space\mathcal{N}_{\bm E,\bm{k}'}^{qn} \space\delta(\varepsilon_{\bm{k}'}^{n} - \varepsilon_{\bm{k}}^{p})  \\
 - \langle U_{\bm{kk}'}^{nq}U_{\bm{k}'\bm{k}}^{np}\rangle\space\mathcal{N}^{qn}_{\bm E,\bm{k}'}\space\delta(\varepsilon_{\bm{k}'}^{n} - \varepsilon_{\bm{k}}^{q}) 
 -\langle U_{\bm{kk}'}^{qn}U_{\bm{k}'\bm{k}}^{np}\rangle\space\mathcal{N}^{nq}_{\bm E,\bm{k}}\space \delta(\varepsilon_{\bm{k}'}^{q} - \varepsilon_{\bm{k}}^{n}) 
\bigg\}.
\end{align}
Further, in the case of the two band model, the scattering term becomes
\begin{align} \nonumber
&\mathcal{J}^{+-}[\mathcal{N}_{\bm E,\bm{k}}] \\
&= \frac{\pi n_i}{\hbar} \sum_{\bm{k}'} \langle U_{\bm{k}\bm{k}'}^{++} U_{\bm{k}'\bm{k}}^{+-} \rangle 
\left\{
\left[ \mathcal{N}_{\bm E,\bm{k}}^{++} -\mathcal{N}_{\bm E,\bm{k}'}^{++} \right] \delta(\varepsilon_{\bm{k}}^{+} - \varepsilon_{\bm{k}'}^{+})
+ \left[ \mathcal{N}_{\bm E,\bm{k}}^{--} -\mathcal{N}_{\bm E,\bm{k}'}^{++} \right]  \delta(\varepsilon_{\bm{k}}^{-} - \varepsilon_{\bm{k}'}^{+})
\right\} \notag \\
& + \frac{\pi n_i}{\hbar} \sum_{\bm{k}'} \langle U_{\bm{k}\bm{k}'}^{+-} U_{\bm{k}'\bm{k}}^{--} \rangle 
\left\{
\left[ \mathcal{N}_{\bm E,\bm{k}}^{++} -\mathcal{N}_{\bm E,\bm{k}'}^{--} \right] \delta(\varepsilon_{\bm{k}}^{+} - \varepsilon_{\bm{k}'}^{-})
+\left[ \mathcal{N}_{\bm E,\bm{k}}^{--} -\mathcal{N}_{\bm E,\bm{k}'}^{--} \right]  \delta(\varepsilon_{\bm{k}}^{-} - \varepsilon_{\bm{k}'}^{-})
\right\}.
\end{align}

For the case of DNLSMs, the products of the disorder matrix elements are 
\begin{align} \nonumber 
U_{\bm{k}\bm{k}'}^{++} U_{\bm{k}'\bm{k}}^{+-} = -\frac{U_0^2}{2 \varepsilon_{\bm k} \varepsilon_{\bm k'}} \sqrt{ \varepsilon^2_{\bm k'} - M^2} \left( i  \varepsilon_{\bm k} \sin \gamma + M \cos \gamma \right)\\
U_{\bm{k}\bm{k}'}^{+-} U_{\bm{k}'\bm{k}}^{--} = +\frac{U_0^2}{2 \varepsilon_{\bm k} \varepsilon_{\bm k'}} \sqrt{ \varepsilon^2_{\bm k'} - M^2} \left( i  \varepsilon_{\bm k} \sin \gamma + M \cos \gamma \right),
\end{align}
where $\gamma = \theta_{\bm k'}-\theta_{\bm k}$. Using the above relation and performing simplifications, we obtain
\begin{align}\label{J_n^+-} \nonumber
&\mathcal{J}^{+-}[\mathcal{N}_{\bm E,\bm{k}}]=\frac{n_i\pi}{ \hbar}\\  
& \times \sum_{\bm{k}'}\langle U_{\bm{k}\bm{k}'}^{++} U_{\bm{k}'\bm{k}}^{+-} \rangle  \left\{
\mathcal{N}_{E,\bm{k}'}^{--} \left[ \delta(\varepsilon_{\bm{k}}^{+} - \varepsilon_{\bm{k}'}^{-}) + \delta(\varepsilon_{\bm{k}}^{-} - \varepsilon_{\bm{k}'}^{-}) \right]
- \mathcal{N}_{E,\bm{k}'}^{++} \left[ \delta(\varepsilon_{\bm{k}}^{+} - \varepsilon_{\bm{k}'}^{+}) + \delta(\varepsilon_{\bm{k}}^{-} - \varepsilon_{\bm{k}'}^{+}) \right]
\right\}.
\end{align}
Further the expression for diagonal part of the density matrix $
\mathcal{N}_{\bm{E}, \bm{k}}^{mm}$ by considering electric field along $x$ direction is $\mathcal{N}_{\bm{E},\bm{k}}^{nn} = \frac{e E_x}{(\tilde{g}_{\bm{k}} + i\tilde{\omega} )} \frac{\partial f^0_{\bm{k}}}{\partial k_x}$, where 
the intraband relaxation time $\tilde{g}_{\bm k}= n_i U_0^2 \int d\bm{k}' \, \delta(\varepsilon_{\bm{k}'}^n - \varepsilon_{\bm{k}}^n) = \frac{n_i U_0^2}{8 \gamma \varepsilon_{\bm{k}}}$.
To do further calculations, let us consider the first part of the Eq.~\eqref{J_n^+-}
\begin{align}\nonumber
\mathcal{J}^{+-}_{I}[\mathcal{N}_{\bm E,\bm{k}}]  
= &\frac{n_i\pi U_0^2 e E_x}{ 2\hbar} v_t \sum_{\bm{k}'}\frac{\sqrt{ \varepsilon^2_{\bm {k}'} - M^2}}{2 \varepsilon_{\bm k} \varepsilon_{\bm {k}'}}  \left( i  \varepsilon_{\bm k} \sin \gamma + M \cos \gamma \right)\\
&\times \left\{
\frac{\delta(\varepsilon_{\bm k'}^{-}-\mu)}{g_{\bm k}+ i \omega}\left[ \delta(\varepsilon_{\bm{k}}^{+} - \varepsilon_{\bm{k}'}^{-}) + \delta(\varepsilon_{\bm{k}}^{-} - \varepsilon_{\bm{k}'}^{-}) \right]\right\}.
\end{align}
Here, we use the partial derivative of the Fermi-Dirac distribution function over the wave vector by writing $\frac{\partial f^0_{\bm{k}}}{\partial \bm{k}} = \frac{\partial f^0_{\bm{k}}}{\partial \varepsilon^{n}_{\bm{k}}} \frac{\partial \varepsilon^{n}_{\bm{k}}}{\partial \bm{k}}$, where $\frac{\partial \varepsilon^{n}_{\bm{k}}}{\partial \bm{k}_x} = \tilde{v}_t + n \frac{\tilde{k}_x (\tilde{\mathcal{K}} - 1)}{\tilde{\mathcal{K}} \varepsilon_{\bm{k}}}$, expressing the Fermi Dirac distribution function within the low temperature regime and expanding it for small tilt $ \frac{\partial f^0_{\bm{k}}}{\partial \varepsilon^{n}_{\bm{k}}} \approx -\delta(\varepsilon^n_{\bm k}-\mu)$. Clearly, the result is nonzero only for finite tilt where other contributions vanishes due to the odd integrand in $\bm k$-space. Similarly, one can get the tilt dependence from other parts of the scattering term. This derives the Eq.~\eqref{eqn:ext_ij}.

\subsection{Derivation of Eq.(12)} \label{Appendix:A3}
Corresponding to the intrinsic part of $S^{np}_{\bm E, \bm k}$, the derivation of the intrinsic conductivity using the definition $\bm{j} = -e \text{Tr}[\bm{v} S_{E, \bm k}]=\sigma \bm {E}$ gives
\begin{align}\label{eqn:Aint_zx}
\sigma_{zx}^{\text{Int}} =-\frac{e}{E_x} \sum_{\bm k}\sum_{n\neq p} v^{pn}_z \frac{{D_{\bm E,{\bm{k}}}^{np}}}{g+i \big(\omega^{pn}+ \omega\big)}.
\end{align}
Here the group velocity is $ \space v^{pn}_z=\space\delta_{pn}\space\partial_{k_z}\varepsilon_{\bm{{k}}}^n+i\space\mathcal{R}^{pn}_{ k_z}\space\omega^{pn}$, where $\hbar\omega^ {pn}= \varepsilon_{{\bm k}}^{p}-\varepsilon_{{\bm k}}^{n}$ is the energy difference between the $p$ and $n$ bands is given below. In the velocity expression, the first term represents the intraband velocity, and the second term shows the interband velocity. Further, to obtain the field driven term $D_{\bm E,{\bm{k}}}^{np}$, we substitute $\mathcal{H}_{\bm E}= e \bm E \cdot \hat{r}$ into the commutator $[\mathcal{H}_{\bm E}, \rho_0]^{np}$, and use $\ket{n,\bm k}=e^{-i \bm k \cdot \bm r}\ket{u_{\bm k}^n}$ to re-express terms of the form $\hat{\bm r}\ket{n,\bm k}=\big[i\frac{\partial}{\partial \bm k} e^{-i \bm k \cdot \bm r}\big]\ket{u_{\bm k}^{n}}$. With this substitution, the driving term can immediately be written in the form 
\begin{align}
    -\frac{i}{\hbar} \bra{n, \bm{k}}[\mathcal{H}_{\bm{E}}, \rho_{0}]\ket{p, \bm{k}} 
    = \frac{e \bm{E}}{\hbar} \cdot \left\{ 
        \delta^{np} \frac{\partial f_0 (\varepsilon_{\bm{k}}^n)}{\partial \bm{k}} 
        + i \mathcal{R}^{np}_{\bm{k}} F^{np} 
    \right\}
\end{align}
Here$\mathcal{R}^{np}_{\bm k}= \braket{u^n_{\bm k}|i \frac{\partial u^n_{\bm k}}{\partial \bm k}}$ represents the Berry connection and $F^{np} = [f_0(\varepsilon^n_{\bm k}) - f_0(\varepsilon^n_{\bm p})] $ is the Fermi occupation number difference. 
Hence, by using the definition of conductivity we can write the intrinsic part as 
\begin{align}\label{eqn:Aint_zx}
\sigma_{zx}^{\text{Int}} =\frac{e^2}{\hbar} \sum_{n\neq p}\sum_{{\bm k}}  \frac{\mathcal{\tilde{Q}}_{zx}^{np} \tilde{F}^{np} \tilde{\omega}^{pn}}{\tilde{g}+\space i \space ( \tilde{\omega}^{pn} + \tilde{\omega})}.
\end{align}
Here we express all factors in normalized form and the term  $\mathcal{\tilde{Q}}_{zx}^{np}= \mathcal{\tilde{R}}^{pn}_{{k}_{z}}\mathcal{\tilde{R}}^{np}_{{k}_{x}} $ is the band-resolved quantum geometric tensor (QGT).
In the case of DNLSMs, the tilt dependence originates only from the occupation probability. Here, at the low temperature the difference between the distribution function associated with different bands $\tilde{F}^{np}$ reduces to the difference between the step function $\Theta(\tilde{\varepsilon}^n_{\bm k}-\tilde{\mu})-\Theta(\tilde{\varepsilon}^p_{\bm k}-\tilde{\mu})$. This results in the integration limit for the angular part as $\phi_{\pm}= \cos^{-1}\bigg[\frac{(\tilde{\mu} \pm \tilde{\varepsilon}_{{\bm k}})}{\tilde{v}_t \mathcal{\tilde{K}}}\bigg]$. 

\twocolumngrid
\bibliography{Ref}

\end{document}